\documentclass[aps,twocolumn,longbibliography]{revtex4-1}
\usepackage{bm,color,amsmath,amssymb,mathrsfs,latexsym,graphicx,psfrag}

\usepackage{soul}

\newcommand{\im}{\mathrm{Im}}

\newcommand{\bk}{{\mathbf k}}

\newcommand{\bq}{{\mathbf q}}

\newcommand{\be}{\begin{equation}}
\newcommand{\ee}{\end{equation}}

\newcommand{\Pf}{\mathrm{Pf}}
\newcommand{\btX}{\bar{\mathrm{X}}}
\newcommand{\btM}{\bar{\mathrm{M}}}
\newcommand{\btD}{\bar{\mathrm{D}}}
\newcommand{\btY}{\bar{\mathrm{Y}}}
\newcommand{\beqq}{\begin{eqnarray*}}
\newcommand{\eeqq}{\end{eqnarray*}}

\def\bea{\begin{eqnarray}}
\def\eea{\end{eqnarray}}
\usepackage{tikz}

\begin{document}
\title{Topological semimetals with Riemann surface states}
\author{Chen Fang$^{\ast\dag{}1,2}$, Ling Lu$^{\dag{}1}$, Junwei Liu$^1$ and Liang Fu$^{\ast{}1}$}
\affiliation{$^1$Department of physics, Massachusetts Institute of Technology, Cambridge, MA 02139, USA}
\affiliation{$^2$Beijing National Laboratory for Condensed Matter Physics, and Institute of Physics, Chinese Academy of Sciences, Beijing 100190, China}

\date{\today}
\begin{abstract}

Riemann surfaces are geometric constructions in complex analysis that may represent multi-valued holomorphic functions using multiple sheets of the complex plane. We show that the energy dispersion of surface states in topological semimetals can be represented by Riemann surfaces generated by holomorphic functions in the two-dimensional momentum space, whose constant height contours correspond  to Fermi arcs. This correspondence is demonstrated in the recently discovered Weyl semimetals and leads us to predict new types of topological semimetals, whose surface states are represented by double- and quad-helicoid Riemann surfaces. The intersection of multiple helicoids, or the branch cut of the generating function, appears on high-symmetry lines in the surface Brillouin zone, where surface states are guaranteed to be doubly degenerate by a glide reflection symmetry. We predict the heterostructure superlattice [(SrIrO$_3$)$_2$(CaIrO$_3$)$_2$] to be a topological semimetal with double-helicoid Riemann surface states.

\end{abstract}
\email{fangc@mit.edu; liangfu@mit.edu\\
$^{\dagger}$The first two authors contributed equally to this work.}
\maketitle

\textbf{Introduction} The study of topological semimetals\cite{Murakami2007} has seen rapid progress since the theoretical proposal of a three-dimensional Weyl semimetal in a magnetic phase of pyrochlore iridates\cite{Wan2011}. In general, topological semimetals are materials where the conduction and the valence bands cross in the Brillouin zone and the crossing cannot be removed by perturbations preserving certain crystalline symmetry such as the lattice translation. Bloch states in the vicinity of the band crossing   possess a nonzero topological index, e.g., the Chern number in case of Weyl semimetals. The nontrivial topology gives rise to anomalous bulk properties of topological semimetals such as the chiral anomaly\cite{Hosur2012,Son2013,Liu2013}. Several classes of topological semimetals have been theoretically proposed so far, including Weyl,\cite{Wan2011,Burkov2011,Burkov2011a,Xu2011,Fang2012,Lu2013,Weng2015,Huang2015,Soluyanov2015}, Dirac\cite{Young2012,Wang2012,Wang2013,Zeng2015} and nodal line semimetals\cite{Burkov2011a,Chiu2014,Phillips2014,Mullen2015,Weng2015a,Xie2015,Zeng2015,Kim2015,Yu2015,Rhim2015,Chiu2015,Carter2012,Chen2015,Fang2015a,Rau2015}, some among which have been experimentally observed\cite{Lu2015,Xu2015a,Lv2015,Shekhar2015,Lv2015a,Yang2015,Xu2015b,Zhang2015,Huang2015a,Liu2014,Liu2014a,Neupane2014,He2014,Jeon2014,Xu2015,Xiong2015,Bian2015}.

Surface states of topological semimetals have attracted much attention. On the surface of a Weyl semimetal, the Fermi surface consists of open arcs connecting the projection of bulk Weyl points onto the surface Brillouin zone \cite{Wan2011}, instead of closed loops. The presence of Fermi arcs on the surface is a remarkable property that directly reflects the nontrivial topology of the bulk, and plays a key role in the experimental identification of Weyl semimetals\cite{Xu2015a,Lv2015,Yang2015}. In contrast, as shown by recent theoretical works\cite{Potter2014,Kargarian2015,Chiu2014,Weng2015a,Kim2015,Yu2015,Chen2015}, existing Dirac and nodal line semimetals do {\it not} have robust Fermi arcs that are stable against symmetry-allowed perturbations. Therefore, the general condition for protected Fermi arcs and their existence in topological semimetals beyond Weyl remain open questions.

In this work, we report the discovery of a new topological semimetal phase in a wide variety of nonsymmorphic crystal structures with the glide reflection symmetry, a combination of a reflection and a translation. Such nonsymmorphic topological semimetals have either Dirac points or Weyl dipoles in the bulk, which are associated with a $Z_2$ topological invariant that we define.
These band crossing points are pairwise connected by symmetry-protected Fermi arcs on the surface, with a unique connectivity determined by the $Z_2$ topological charge.
Interestingly, these surface states have a momentum-energy dispersion that can be mapped to the Riemann surface for a holomorphic function whose singularity corresponds to the bulk Dirac or Weyl points, and hence dubbed ``Riemann surface states''. By relating the $Z_2$ topological index to rotation eigenvalues of energy bands, we provide a simple criterion for the nonsymmorphic topological semimetal phase
and predict its material realization in the recently synthesized superlattice heterostructure of iridates\cite{Matsuno2015} [(SrIrO$_3$)$_{2m}$(CaIrO$_3$)$_{2n}$].

\begin{figure}[tbp]
\includegraphics[width=0.5\textwidth]{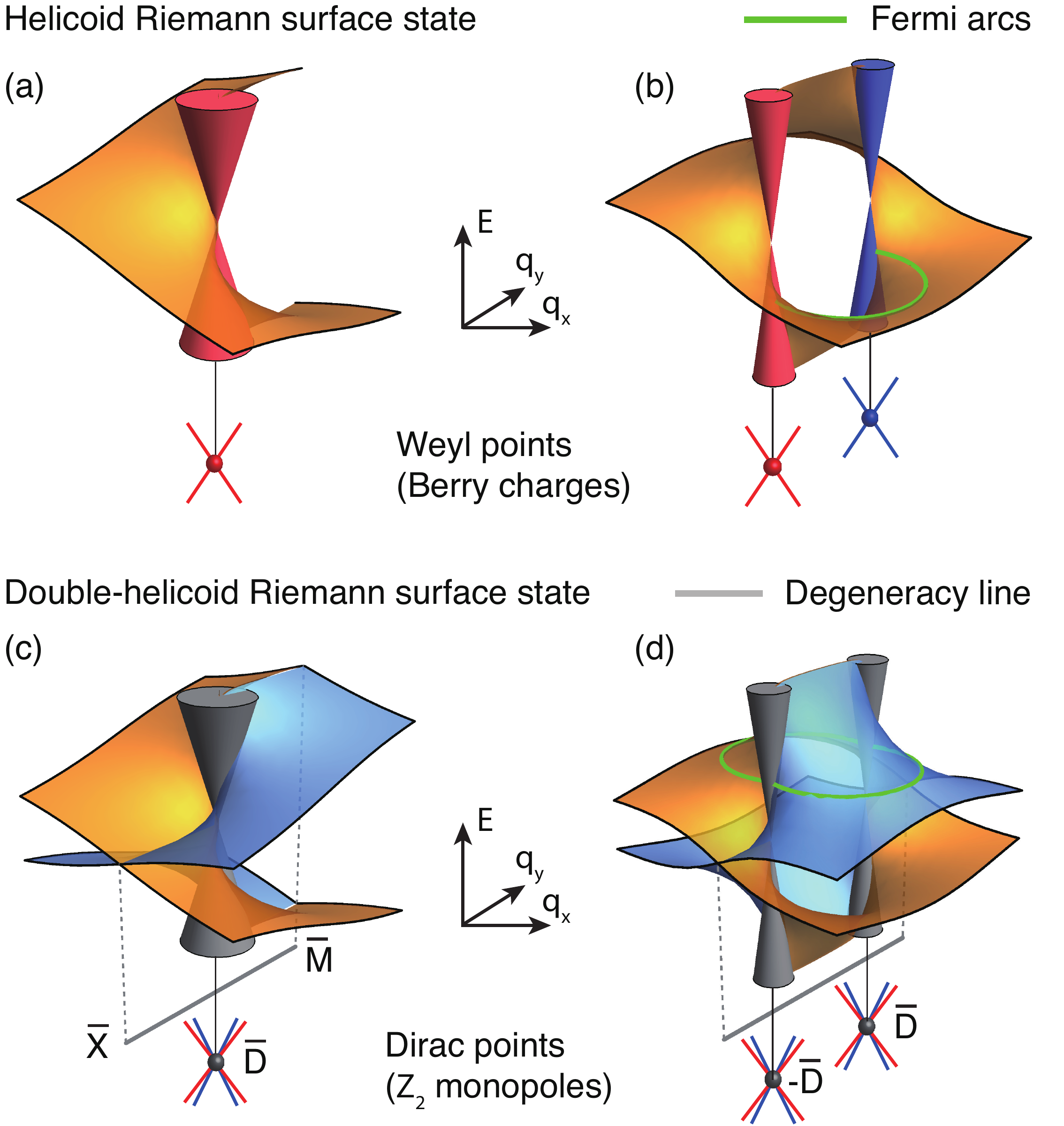}
\caption{(a) The surface dispersion near the projection of a Weyl point with Chern number $+1$, where the red solid cones are the projection of the bulk states and the helicoid sheet represents the surface states. This is also the Riemann surface of $\im[\log(q)]$. (b) The surface dispersion near the projections of a pair of Weyl points with opposite Chern numbers, where the red and the blue cones represent the bulk states projection, and the green contour is an iso-energy contour also known as a Fermi arc. This is also the Riamann surface of $\im[\log\frac{q-k_1}{q-k_2}]$. (c) The surface dispersion of the nonsymmorphic Dirac semimetal near the Dirac point, where the gray cones represent the projection of bulk states. This is also the Riemann surface of the function given in Eq.(\ref{eq:Riemann2}) in the text. (d) The surface dispersion near two nonsymmorphic Dirac points, with iso-energy contours of two Fermi arcs.}
\label{fig:1}
\end{figure}

\textbf{Weyl semimetal and Riemann surface} We start by considering the energy-momentum relation $E(\bk_\parallel)$ of surface states of Weyl semimetals, where $\bk_\parallel$ is the surface momentum. $E(\bk_\parallel)$ is bounded by the conduction and valence band edges in the bulk cones [see the head-to-head cones in Fig.\ref{fig:1}(a)],
obtained by collapsing energies of bulk states with the same $\bk_\parallel$ at different perpendicular momenta $k_z$. In the most generic case, we assume that there be $N_s$ surface bands, $E_1(\bk_\parallel)<E_2(\bk_\parallel)<...<E_{N_s}(\bk_\parallel)$. Consider a loop in the surface Brillouin zone enclosing the projection of the Weyl point. The Chern number of the Weyl point dictates that\cite{Wan2011}, as a $\bk$-point moves one round along the loop counterclockwise (clockwise), the energy of the state does not return to the same value, but moves one band higher (lower), that is, $E_n(\bk_\parallel)\rightarrow{E}_{n+1}(\bk_\parallel)$ [$E_n(\bk_\parallel)\rightarrow{E}_{n-1}(\bk_\parallel)$]. As $\bk_\parallel$ keeps circling the loop counterclockwise (clockwise), the band index keeps increasing (decreasing) before the state merges into the conduction (valence) bulk. In this process, the dispersion along the loop maps out a spiral that connects the two bulk cones, and as one sweeps the radius of the loop, the spirals at different radii form a helicoid as shown in Fig.\ref{fig:1}(a).

The winding of the energy dispersion along any loop enclosing the Weyl point is the same as the winding of the phase of a holomorphic function along any loop enclosing a simple~(linear order) zero. Near a simple zero, a general holomorphic function takes the form $f(z)=z-z_0+O[(z-z_0)^2]$ up to an overall factor. As $z$ goes around $z_0$ counterclockwise (clockwise), the phase of $f(z)$ increases (decreases) by $2\pi$. Therefore, the phase of $f(z)$ near $z_0$, or the imaginary part of $\log[f(z)]$, is topologically equivalent to the dispersion of the surface states near the projection of a positive Weyl point. Similarly, one can show that the phase of a holomorphic function near a simple pole is equivalent to the energy dispersion near the projection of a negative Weyl point. This topological equivalence can be expressed as
\bea\label{eq:Weyl}
E(\bq_\parallel)\sim\im[\log(q^{\pm1})],
\eea
where $\bq_\parallel$ is the surface momentum relative to the Weyl point projection and $q=q_a+iq_b$, and $\pm1$ corresponds to Weyl point of positive and negative monopole charge. There is one caveat in understanding Eq.(\ref{eq:Weyl}): while the function on the right-hand-side ranges from negative to positive infinity, the energy of the surface bands always merge into the bulk. This infinite winding of the surface dispersion implies that the theory cannot be made ultraviolet-complete in 2D, but is only consistent for the surface states of some topologically nontrivial 3D bulk: a demonstration of the bulk-edge correspondence principle in Weyl semimetals.

In complex analysis, the plot of the real or the imaginary part of a multi-valued holomorphic (meromorphic) function is called a Riemann surface, which is a surface-like configuration that covers the complex plane a finite (compact) or infinite (noncompact) number of times\cite{Knopp1996}. Eq.(\ref{eq:Weyl}) establishes the topological equivalence between the surface dispersion of a Weyl semimetal and a \emph{noncompact} Riemann surface. Both share the following characteristic feature: There is no equal energy (equal height) contour that is both closed and encloses the projection of the Weyl point, a feature that directly leads to the phenomenon of ``Fermi arcs''. This topological equivalence can be extended to the case of multiple Weyl points. If there are two Weyl points' projections at $(k_{1a},k_{1b})$ and $(k_{2a},k_{2b})$, then the corresponding function is simply $\log[(q-k_1)(q-k_2)^{-1}]$, where $k_i=k_{ia}+ik_{ib}$, whose imaginary part is plotted in Fig.\ref{fig:1}(b). Cutting the dispersion at any energy, the iso-energy contour is an arc connecting $\bk_{1}$ and $\bk_{2}$.
\begin{figure}[tbp]
\includegraphics[width=0.5\textwidth]{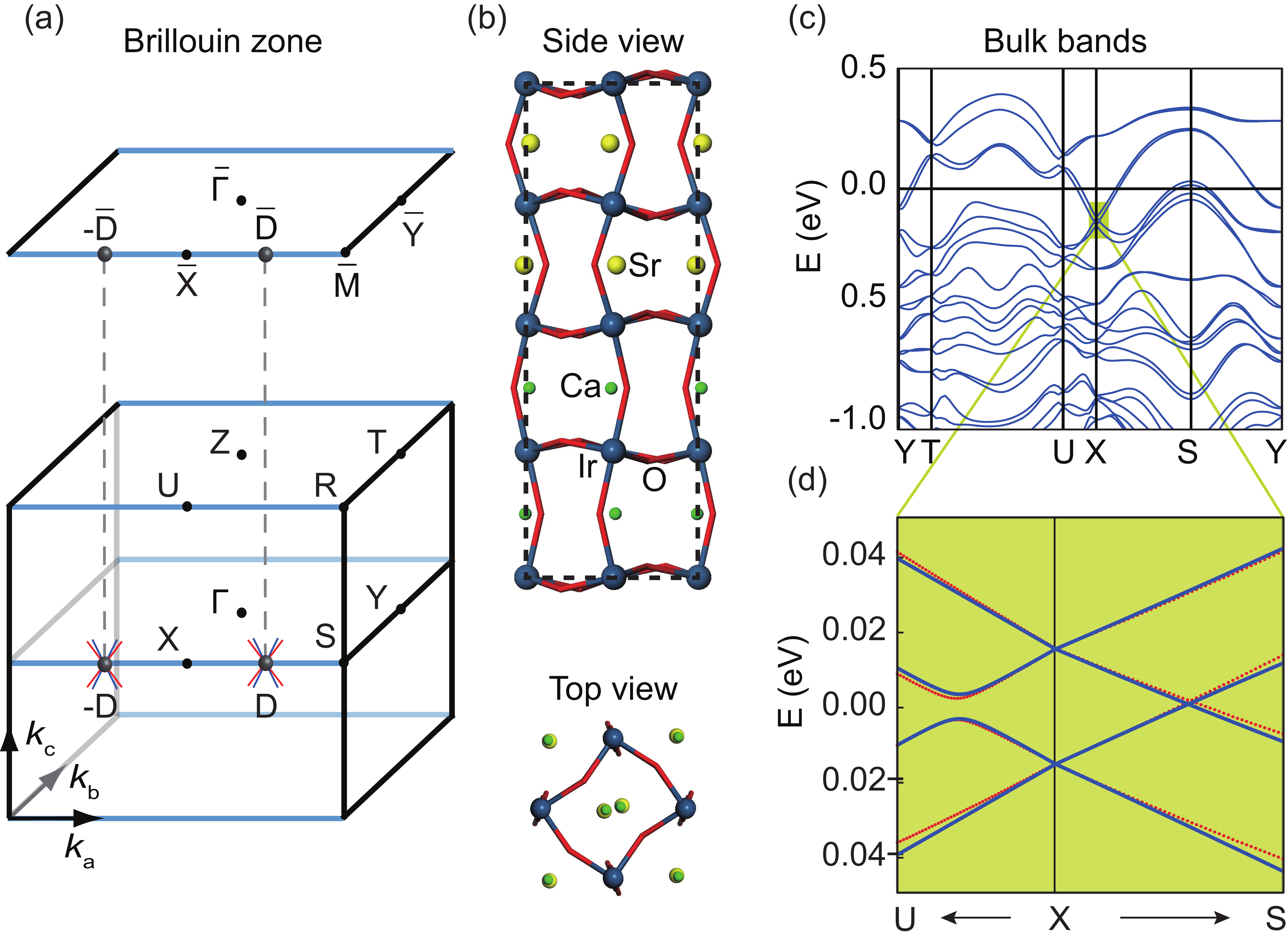}
\caption{(a) The Brillouin zone and the $(001)$-slab surface Brillouin zone of a orthorhombic lattice with a glide reflection, where the lines of double degeneracy are marked in blue and the Dirac points and their projections are marked by green dots. (b) One unit cell of heterostructure superlattice [(SrIrO$_3$)$_{2m}$(CaIrO$_3$)$_{2n}$] with $m=n=1$. (c) The bulk band structure of the superlattice along the path YTUXSY, calculated from first principles. (d) The zoomed-in band structure of the same system near X, where the first principles results (red dots) are fitted using a tight-binding model modified from Ref.[\onlinecite{Carter2012}] (blue line).}
\label{fig:2}
\end{figure}

\textbf{Double-helicoid Riemann surface state protected by one glide reflection symmetry in a Dirac semimetal}
A Dirac point can be considered as the superposition of two Weyl points with opposite Chern numbers\cite{Young2012,Wang2012}, the same way the 3D massless Dirac equations decouple into two sets of Weyl equations\cite{Peskin1995}. The surface states near the projection of a Dirac point is hence a superposition of a helicoid and an anti-helicoid Riemann surface as shown in Fig.\ref{fig:1}(c), which cross each other along certain lines, and may have two Fermi arcs\cite{Wang2012,Wang2013,Xu2015}. Yet, if there be no additional symmetry that protects their crossing, hybridization along the crossing lines opens a gap and the double-helicoid structure of the surface dispersion is lost and the Fermi arcs also disappear. This has been the case of all Dirac semimetals discovered so far. Below we show that a nonsymmorphic symmetry\cite{Parameswaran2013,Freed2013,Liu2014b,Fang2015,Shiozaki2015,Fang2015a,Varjas2015,Watanabe2015,lu2015three,Wang2015} protects the crossing and with it the topological surface states.

Consider a three-dimensional system with the following symmetries: a glide reflection, $G$, that reverses the $a$-direction then translates by half lattice constant along the $b$-direction, and time-reversal symmetry, $T$. Define the antiunitary symmetry $\Theta$ as their composition
\bea\label{eq:Thetadef}
\Theta\equiv{G}*T:\;(x,y,z,t)\rightarrow(-x,y+1/2,z,-t),
\eea
where $(x,y,z)$ are the spatial coordinates along $a,b,c$-axes in unit of the corresponding lattice constants.
Eq.(\ref{eq:Thetadef}) implies that the momentum of a single quasiparticle, $(k_a,k_b,k_c)$, is sent to $(k_a,-k_b,-k_c)$. Importantly, the square of $\Theta$
\bea\label{eq:Theta2}
\Theta^2=G^2T^2=T_{010}=e^{-ik_b},
\eea
where $T_{010}$ is the unit lattice translation along the $b$-direction. Specially at the Brillouin zone boundary $k_b=\pi$, we have $\Theta^2=-1$. This leads to double degeneracy of all states on two high-symmetry lines, UR and XS, analogous to the well-known Kramers' degeneracy\cite{Kramers1930} [blue lines in the 3D Brillouin zone of Fig.~\ref{fig:2}(a)], with the key difference that the latter leads to double degeneracy at high-symmetry points in a spinful system, $\Theta$ leads to double degeneracy along the whole high-symmetry lines in both spinful and spinless systems.

Then we consider the states on the $(001)$-surface. In the surface Brillouin zone, Eq.(\ref{eq:Theta2}) leads to double degeneracy along $\btX\btM$ [blue line in the surface Brillouin zone of Fig.~\ref{fig:2}(a)]. This degeneracy is exactly what is needed to protect the double-helicoid surface states shown in Fig.\ref{fig:1}(c): if there is a projection of Dirac point on $\btX\btM$ and the two helicoids intersect along $\btX\btM$, the symmetry guaranteed double degeneracy disallows their hybridization. In the double-helicoid dispersion, each iso-energy contour must contain two arcs emanating from the projection of the Dirac point. Due to time-reversal, each projection of Dirac point at $\bar{D}$ is accompanied by one at $-\bar{D}$. The surface dispersion with two Dirac points is shown in Fig.\ref{fig:1}(d), and each iso-energy contour must contain two arcs connecting $\bar{D}$ and $-\bar{D}$.

As the surface dispersion near a Weyl point projection can be mapped to the Riemann surface of $\log(z)$, a natural question is if the the surface dispersion of the Dirac semimetals can also be mapped to some Riemann surface representing a holomorphic function. Configuration of two surfaces crossing along certain lines reminds us of Riemann surfaces of holomorphic functions involving a fractional power. For example, $f(z)=\sqrt{z^2}$ has two branches $f_\pm(z)=\pm{z}$, and the imaginary parts of the two branches meet each other at the real axis, as $\im(z)=\im(-z)=0$ for $z\in\mathbb{R}$. Since the dispersion near the positive and the negative Weyl points are mapped to the phases of $z$ and $z^{-1}$, what we are looking for is a homomorphic function whose two branches are $\log{z}$ and $\log{z^{-1}}$. These considerations suggest the following choice
\bea\label{eq:Riemann2}
E(\bq_\parallel)\sim\im[\log({q}+q^{-1}+\sqrt{q^2+q^{-2}-2})],
\eea
where $\bq=\bk-\btD$.

According to the bulk-edge correspondence principle, the nontrivial surface state protected by $\Theta$ suggests a nontrivial bulk topology near each Dirac point. In the main text, for concision, we only make the following remarks and leave the detailed discussion of bulk topology to the Supplemental Materials: (i) A Dirac point is either on XS or UR; (ii) On a sphere enclosing the Dirac point, there is a $Z_2$ topological invariant protected by $\Theta$; (iii) if inversion is also present and if the system is spinful (with SOC), the invariant can be expressed in terms of rotation eigenvalues of bands along $XS$ or $UR$, analogous to the Fu-Kane formula for topological insulators\cite{Fu2007}. In the absence of additional symmetry but only $\Theta$, the Dirac point is not protected and is split into two Weyl points of opposite charge along $XS$ or $UR$, termed a Weyl dipoles, related to each other by $\Theta$. In this case, we consider a sphere enclosing the Weyl dipoles. The Chern number of the sphere is zero due to the cancelation of monopole charge, but the new $Z_2$ topological charge is nontrivial. In this case, on the surface, Fermi arcs only connect Weyl points from \emph{different} Weyl dipoles, and the two Weyl points within one Weyl dipole are not connected by any Fermi arc.

\textbf{Material realization in iridates}
\begin{figure}[tbp]
\includegraphics[width=0.45\textwidth]{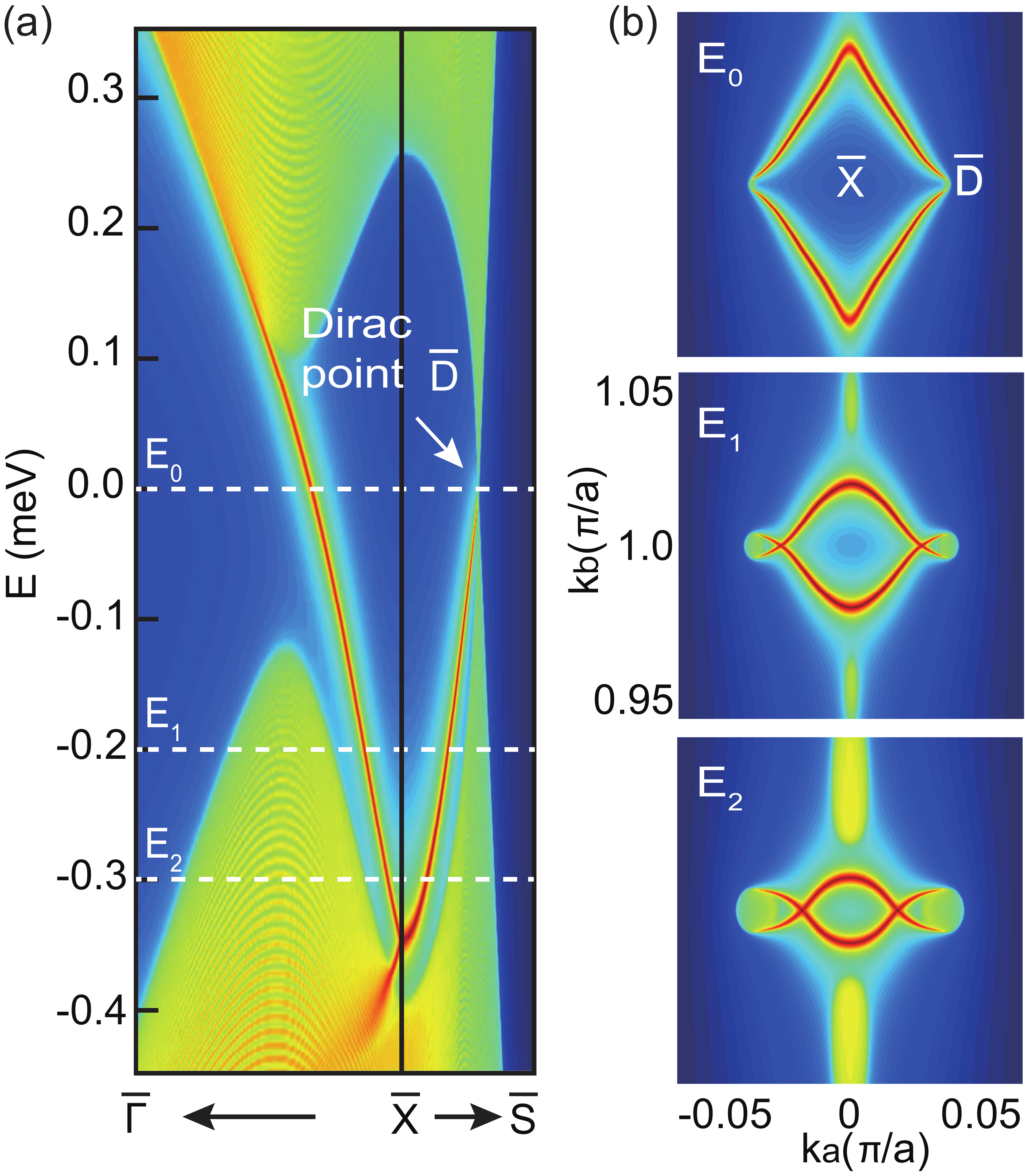}
\caption{(a) The spectral weight of the top surface of a $(001)$-slab, along the path $\bar\Gamma$-$\btX$-$\btM$, calculated from the tight-binding model used in fitting. (b) The spectral weight of the $(001)$-surface at, from top to bottom, $E_{0,1,2}$ respectively.}
\label{fig:3}
\end{figure}
Perovskite iridate SrIrO$_3$ was shown to be a TSM with a degenerate nodal line protected by a twofold screw axis\cite{Carter2012,Fang2015a}. It was found in Ref.[\onlinecite{Carter2012}] that if a staggering chemical potential propagating along the $[001]$-direction, the nodal line is gapped at all but two points. Based on this finding, we propose to realize the nonsymmorphic Dirac semimetal in a [(SrIrO$_3$)$_{2m}$(CaIrO$_3$)$_{2n}$] superlattice heterostructure shown in Fig.\ref{fig:2}(b). For $m=n=1$, we perform a first principles calculation for the bulk band structure, and find a pair of Dirac points along XS that are close and symmetric on each side of X shown in Fig.\ref{fig:2}(c). We modify the tight-binding model given in Ref.[\onlinecite{Carter2012}] such that its band structure quantitatively matches that from the first principles calculation near X [Fig.\ref{fig:2}(d)]. (Also find details of the tight-binding model in Supplementary Materials.) Using the fitted tight-binding model, we calculated the spectral weight of the states near the top surface of a $(001)$-slab, along high-symmetry lines in surface Brillouin zone [Fig.\ref{fig:3}(a)], and the 2D surface Brillouin zone near $\btX$ at three different energies [Fig.\ref{fig:3}(b)], where the double Fermi arcs can be seen. We note that as the energy decreases from $E_0$, the energy of the bulk Dirac point, (i) bulk pockets emerge near the projection of the Dirac point and (ii) more importantly, the configuration of the two arcs rotation around the projections of the Dirac points, such that in Fig.\ref{fig:3}(b), the two arcs cross each other along $\btX\btM$, where the crossing point is always protected by $\Theta$.

\begin{figure}[tbp]
\includegraphics[width=0.3\textwidth]{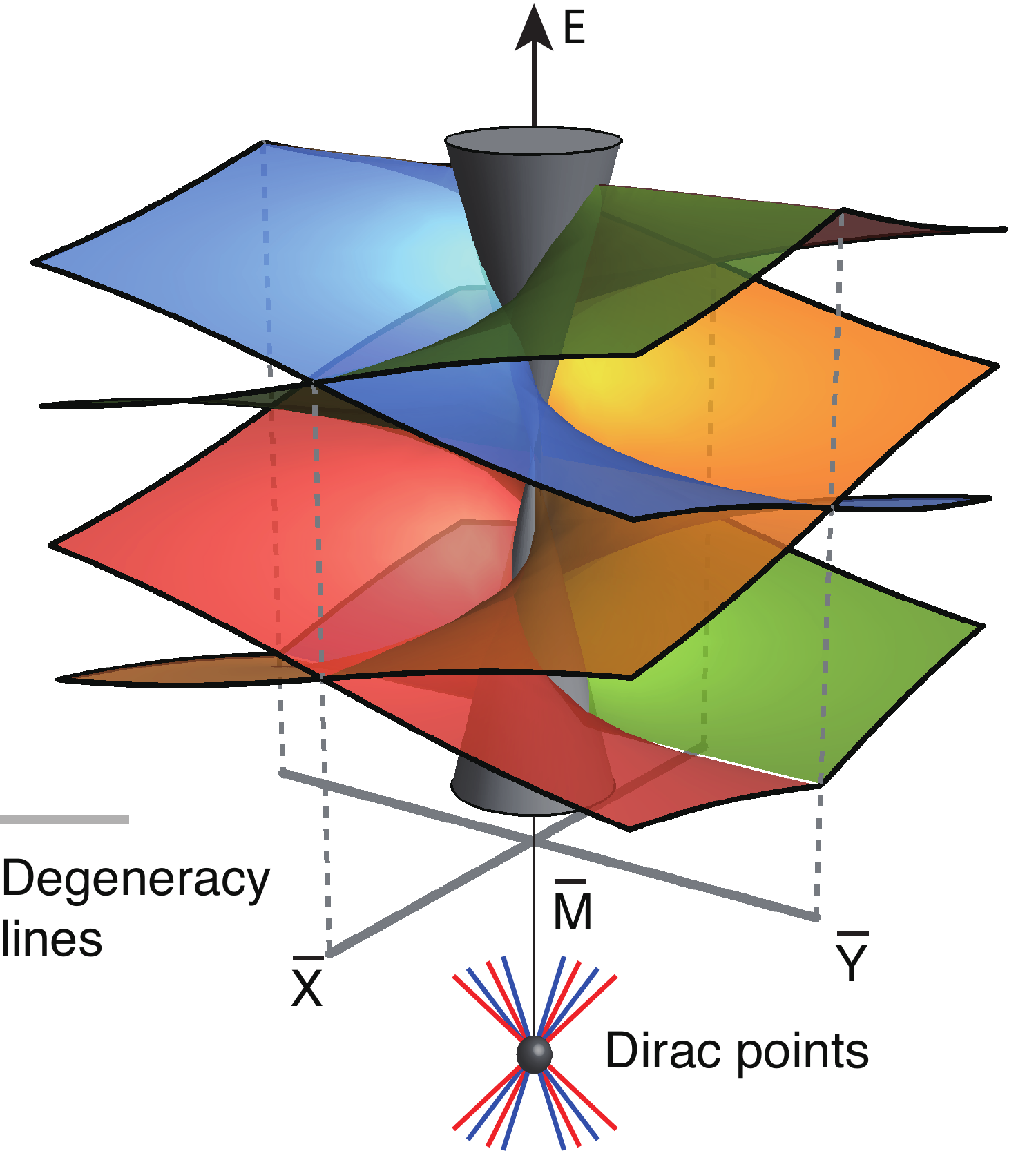}
\caption{The quad-helicoid surface state dispersion, consisting of two helicoids (blue and red) and two anti-helicoids (green and yellow). The blue and the green, and the yellow and the red cross each other along $\btM\btX$, and the blue and the yellow and the red and the green cross each other along $\btM\btY$.}
\label{fig:4}
\end{figure}

\textbf{Quad-helicoid surface state protected by two glide reflections}
Finally, we point out that new types of TSM may exist if additional nonsymmorphic symmetries on the surface are present, with their own characteristic surface dispersions.
As an example, we assume there be an additional glide plane, $G'$, that is perpendicular to $G$,
\bea
G':\;(x,y,z)\rightarrow(x+1/2,-y,z).
\eea
Following similar steps, we find that $\Theta'\equiv{}G'*T$ guarantees double degeneracy along $\btY\btM$, so that if both $\Theta$ and $\Theta'$ are present, all bands are doubly degenerate along $\btX\btM$ and $\btY\btM$. This double degeneracy protects a unique nontrivial surface dispersion consisting of four spiral surfaces near $\btM$, as shown in Fig. \ref{fig:4}, or can be considered as the superposition of the surface dispersions near four Weyl points, two positive and two negative ones. This dispersion has a new type of $Z_2$ spectral flow between two perpendicular lines of $\btX\btM$ and $\btY\btM$: two bands from a degenerate pair at $\btX\btM$ flow to different degenerate pairs at $\btY\btM$. A generic iso-energy contour of this quad-helicoid surface dispersion consists of four Fermi arcs emanating from $\btM$. Since there is only one $\bar{\textrm{M}}$ inside the surface Brillouin zone, we argue that no topological invariant can be defined for the band crossings which project to $\bar{\textrm{M}}$, or the Nielson-Ninomiya theorem would be violated. We conjecture that the system belongs to filling enforced semimetals discussed in Ref.[\onlinecite{Parameswaran2013,Watanabe2015}], where the band crossings are guaranteed by the space group at certain integer fillings.
In this case, the surface dispersion can also be mapped to a Riemann surface. Since the dispersion can be considered as the superposition of four spiral surfaces, we consider a holomorphic function with four branches. $\Theta$ and $\Theta'$ require that two branches meet along $\bar{\textrm{X}}\bar{\textrm{M}}$ (defined as the real axis) and $\bar{\textrm{Y}}\bar{\textrm{M}}$ (defined as the imaginary axis), respectively. The function we choose is
\bea\label{eq:omega}
E(\bq_\parallel)\sim\im[\log(\sqrt{q^2+q^{-2}+2}+\sqrt{q^2+q^{-2}-2})],
\eea
where $q=(k_a-\pi)+i(k_b-\pi)$

\textbf{Conclusions} In this paper we theoretically find two new classes of topological semimetals that have multiple Fermi arcs on the surface protected by nonsymmorphic glide reflections symmetries and time-reversal. We observe that so far, all topological semimetals with protected Fermi arcs have surface dispersions that are topologically equivalent to the noncompact Riemann surfaces representing simple holomorphic functions. We propose superlattice heterostructure [(SrIrO$_3$)$_{2m}$(CaIrO$_3$)$_{2n}$] as a nonsymmorphic Dirac semimetal with two Fermi arcs on the $(001)$-plane protected by one glide reflection.

\textbf{Acknowledgements}
{
We thank Timothy H. Hsieh for discussion. C.F. thanks Yige Chen for helpful discussion on the tight-binding model. C.F. thanks Jian Liu for fruitful discussions on potential material systems.
C.F. and L.F. were supported by S3TEC Solid State Solar Thermal Energy Conversion Center, an Energy Frontier Research Center funded by the U.S. Department of Energy (DOE), Office of Science, Basic Energy Sciences (BES), under Award No. DE-SC0001299/DE-FG02-09ER46577.
L.L. was supported in part by U.S.A.R.O. through the ISN under Contract No. W911NF-13-D-0001, in part by the MRSEC Program of the NSF under Award No. DMR-1419807, and in part by the MIT S3TEC EFRC of DOE under Grant No.DESC0001299.
J. L was supported by the STC Center for Integrated Quantum Materials, NSF Grant No. DMR-1231319.

\clearpage
\begin{appendix}
\section{Bulk invariant protected by $G*T$}

\subsection{$Z_2$ invariant protected by $G*T$}

The bulk invariant is defined on a sphere in the Brillouin zone that encloses some band crossings (either nodal points or nodal lines), and on the surface of that sphere, the conduction and the valence bands have a finite direct gap and hence can be separated. For our case, due to $\Theta$, the generic band crossing is either a pair of opposite Weyl points symmetric about XS or UR. We use a sphere centered at $\bk_0$, a point on XS, with radius $k_r$. Each point on the sphere is parameterized by $(\theta,\phi)$:
\bea
&&(k_a,k_b,k_c)\\
\nonumber&=&(k_{0a}+k_r\cos\theta,k_{0b}+k_r\sin\theta\cos\phi,k_{0c}+k_r\sin\theta\sin\phi).
\eea

The derivation of the $Z_2$ invariant on a sphere invariant under $\Theta=G*T$ closely follows the derivation of the $Z_2$ invariant of 2D topological insulators. (See Ref.[\onlinecite{Fu2007}].)

First we parameterize the sphere such that under $\Theta$, a point at $(\theta,\phi)$ is mapped to $(\theta,\phi+\pi)$. Since the total Chern number on the sphere vanish, and we can hence in principle choose a smooth gauge for all occupied bands, denoted by $|u_{n\in{occ}}(\theta,\phi)\rangle$. We can hence define the following sewing matrix
\bea
W_{mn}(\theta,\phi)=\langle{u}_m(\theta,\phi+\pi)|\Theta|u_n(\theta,\phi)\rangle.
\eea
At $\theta=0,\pi$, we have
\bea
W_{mn}(0/\pi)&=&\langle{u}_m(0/\pi)|\Theta|u_n(0/\pi)\rangle\\
\nonumber
&=&(\langle{\Theta}{u}_m(0/\pi)|\Theta^2|u_n(0/\pi)\rangle)^2\\
\nonumber
&=&-\langle{u}_n(0/\pi)|\Theta|u_m(0/\pi)\rangle\\
\nonumber
&=&-W_{nm}(0/\pi),
\eea
i.e.,
\bea
W=-W^T(0/\pi).
\eea
Hence we can define the following $Z_2$ quantity
\bea\label{eq:PfZ2}
\delta_0=\frac{\Pf[W(0)]}{\sqrt{\det[W(0)]}}\frac{\Pf[W(\pi)]}{\sqrt{\det[W(\pi)]}},
\eea
where $\Pf$ stands for the Pfaffian of an antisymmetric matrix. Eq.(\ref{eq:PfZ2}) defines a $Z_2$ quantity which is either $+1$ or $-1$, because $\Pf^2=\det$ in general.

To prove that the $Z_2$ quantity is also gauge invariant, consider changing the gauge by a smooth unitary $N_{occ}$-by-$N_{occ}$ matrix $U(\theta,\phi)$
\bea
|u'_m(\theta,\phi)\rangle=\sum_n{U}_{mn}(\theta,\phi)|u_n(\theta,\phi)\rangle.
\eea
It is straight forward to see that after the transform, the sewing matrix becomes
\bea
W'(\theta,\phi)=U^T(\theta,\phi+\pi)W(\theta,\phi)U(\theta,\phi),
\eea
so that at $\theta=0,\pi$
\bea\label{eq:UonW}
\Pf[W'(0/\pi)]&=&\det[U(0/\pi)]\Pf[W(0/\pi)],\\
\nonumber
\det[W'(0/\pi)]&=&\det[U^T(0/\pi)]\det[W(0/\pi)]\det[U(0/\pi)]\\
\nonumber&=&\det[W(0/\pi)]{\det}^2[U(0/\pi)].
\eea
Substituting Eqs.(\ref{eq:UonW}) into Eq.(\ref{eq:PfZ2}), we find
\bea
\delta'_0=\delta_0.
\eea

\subsection{Simplification of the $Z_2$-invariant in spinful systems in the presence of inversion}

In this section, we show how the $Z_2$-invariant given in terms of Pfaffians in Eq.(\ref{eq:PfZ2}) simplifies in the presence of inversion symmetry in a spinful system. In this section, we closely follow the derivation of the original Fu-Kane formula in topological insulators with inversion symmetry, which can be briefly summarized as follows: (i) the bands at time-reversal invariant momenta are also eigenstates of the inversion; (ii) each state and its time-reversal partner have the same inversion eigenvalue, so that each Kramers' pair at a TRIM maps to an eigenvalue of either $+1$ or $-1$; (iii) the product of the inversion eigenvalues of all occupied Kramers' pairs at all TRIM is the same as the Pffafian invariant.

In our case, the time-reversal symmetry is replaced by $\Theta=G*T$ and the points that are invariant under $\Theta$ are $\theta=0,\pi$ on the sphere. The inversion symmetry itself is not a good quantum number at these points, but the composition symmetry $R_2\equiv{P*G}$ is. We will now prove that for each degenerate pair of states at $\theta=0,\pi$ have the same eigenvalue of $R_2$.

We distinguish two cases of (i) the inversion center is within the glide plane and (ii) the inversion center is not within the glide plane. A generic inversion operation takes the form
\bea
P:\;(x,y,z)\rightarrow(\frac{\lambda}{2}-x,\frac{\mu}{2}-y,\frac{\nu}{2}-z),
\eea
where $\lambda,\mu,\nu=0,1$. If $\lambda=0$, then the inversion center, $(0,\mu,\nu)/2$ is on the glide plane; if $\lambda=1$, then the inversion center $(1/2,\mu/2,\nu/2)$ is away from the glide plane.

If the inversion center is inside the glide plane, then we have
\bea
R_2:\;(x,\frac{\mu}{2}-y-\frac{1}{2},\frac{\nu}{2}-z),
\eea
and
\bea
R_2^2:(x,y,z).
\eea
Yet, since in spin space $R^2$ is equivalent to a full spin rotation, we have
\bea\label{eq:R2inplane1}
R^2=-1.
\eea
Also, the commutation relation between $R_2$ and $\Theta$ can be shown to be
\bea\label{eq:R2inplane2}
R_2\Theta=T_{010}\Theta{R}_2=e^{-ik_b}\Theta{R}_2.
\eea
From Eq.(\ref{eq:R2inplane1}), each state at $\theta=0,\pi$ is also an eigenstate of $R_2$ with eigenvalue of either $+i$ or $-i$. Using Eq.(\ref{eq:R2inplane2}), we see that for each eigenstate of $R_2$ with eigenvalue $+i$
\bea
R_2\Theta|+i\rangle=e^{-ik_b}\Theta{R}_2|+i\rangle=-e^{-ik_b}i\Theta|+i\rangle,
\eea
i.e., $\Theta|+i\rangle$ is an eigenstate of $R_2$ with eigenvalue $-e^{-ik_b}i=+i$ at $\theta=0,\pi$. Hence the two states in one degenerate pair at $\theta=0,\pi$ have the same eigenvalue of $R_2$.
Following Ref.[\onlinecite{Fu2007}], we show that the $Z_2$ invariant can be expressed in terms of these eigenvalues
\bea
\delta_0=\prod_{n=1,...,N_{occ}/2}\frac{\gamma_{2n}(0)}{\gamma_{2n}(\pi)}.
\eea

If the inversion center is away from the glide plane, we have
\bea
R_2:(x+\frac{1}{2},\frac{\mu}{2}-y-\frac{1}{2},\frac{\nu}{2}-z),
\eea
which is in fact a twofold screw axis. The square of $R_2$ is
\bea\label{eq:R2notinplane1}
R_2^2:(x+1,y,z).
\eea
Again, considering the spin rotation in $R_2$, the eigenvalues are $\pm{i}e^{-ik_a/2}$ from Eq.(\ref{eq:R2notinplane1}).
The commutation relation between $R_2$ and $\Theta$ is
\bea
R_2\Theta=T{110}\Theta{R}_2=e^{-ik_a-ik_b}\Theta{R}_2,
\eea
so that for an eigenstate of $R_2$ with eigenvalue $+ie^{-ik_a/2}$, we have
\bea\label{eq:R2notinplane2}
R_2\Theta|+ie^{-ik_a/2}\rangle=-ie^{-ik_a/2-ik_b}|+ie^{-ik_a/2}\rangle.
\eea
Eq.(\ref{eq:R2notinplane2}) shows that at $\theta=0,\pi$ (where $k_b=\pi$), the two degenerate states have the same eigenvalue of $R_2$. Therefore the following expression
\bea
\delta_0=\prod_{n=1,...,N_{occ}/2}\frac{e^{ik_0/2}\gamma_{2n}(0)}{e^{-ik_0/2}\gamma_{2n}(\pi)}.
\eea

\section{Splitting of the nonsymmorphic Dirac point in the absence of inversion}

In this section, we lift the symmetry of inversion, keeping glide reflection and time-reversal. Without the inversion, the bands are in general singly degenerate, and a single Dirac point splits into two Weyl points. Since glide reflection inverts the monopole charge of the Weyl point and time-reversal preserves it, the configuration of the split Dirac point is such that $W_{1}$ is related by $W_2$ by $\Theta=G*T$, while $W'_{1,2}$ are related to $W_{1,2}$ by time-reversal, in the presence of a inversion breaking perturbation. Yet it is important to note that even in this case, the system is not a generic Weyl semimetal, because each pair of Weyl points related by $G*T$, $W_1$ and $W_2$ for example, carry a $Z_2$ topological charge. Consider a sphere enclosing such a pair, and the definition of the $Z_2$ invariant only depends on the presence of $G*T$. Therefore, if this invariant is nontrivial in the presence of inversion due to the Dirac point, it remains nontrivial after the splitting. This $Z_2$ topological charge has two consequences: (i) on the surface preserving $G$, the Fermi arcs must \emph{not} connect the projections of Weyl points that are related by $G*T$, and (ii) there must be an even number of such pairs of Weyl points due to the Nielson-Ninomiya theorem.

\section{Some details of the numerics}
The band structures of (SrIrO$_3$)$_{2m}$(CaIrO$_3$)$_{2n}$ are calculated in the framework of density functional theory (DFT) including the Hubbard $U$, as implemented in the Vienna \emph{ab initio} simulation package (VASP) \cite{VASP} by using generalized gradient approximation (GGA) of exchange-correlation function in the Perdew-Burke-Ernzerhof (PBE) form \cite{PBE}. The projector augmented wave method \cite{PAW1} was applied to model the core electrons. Monkhorst-Pack \emph{\bf k}-point sampling of 4$\times$4$\times$2 was used for (m=1, n=1). Energy cutoff of the planewave basis was fully tested, and atomic structures were optimized with maximal residual forces smaller than 0.01 eV/{\AA}. Spin-orbit coupling (SOC) was included in all calculations. For the Hubbard $U<2$, all the results are similar, and here we only show the results for $U=0$ for the sake of simplicity.

For SrIrO$_3$ (i.e. m=1, n=0), we got the similar results as the previous study\cite{Carter2012}, with Dirac nodal line around the Fermi energy. For (SrIrO$_3$)$_{2}$(CaIrO$_3$)$_{2}$, the Dirac nodal will fold around $X$ point and is gapped expect a pair of Dirac points along $XS$ since the two-fold screw symmetry is broken. The properties of band structure around the Dirac points of (SrIrO$_3$)$_{2}$(CaIrO$_3$)$_{2}$ can well described by adding some mass terms based on the TB model in Ref. \cite{Carter2012}.
\begin{widetext}
The tight-binding model for (SrIrO$_3$)$_{2}$(CaIrO$_3$)$_{2}$ is
\beqq
H=\left(
          \begin{array}{cccc}
            H_0+H_1 & T+T_1 & 0 & e^{-ik_z}(T-T_1)^\dag  \\
            (T+T_1)^\dag & (H_0-H_1)\epsilon + m_1 & T-T_1 & 0 \\
             0 & (T-T_1)^\dag  & H_0+H_1 & T+T_1  \\
             e^{ik_z}(T-T_1)& 0 & (T+T_1)^\dag  & H_0-H_1 \\
          \end{array}
        \right)
\eeqq
where $H_0=2t_p(\rm{cos} k_x + \rm{cos} k_y)\tau_x$,
$H_1 =(t_{1p} \rm{cos} k_y + t_{2p} \rm{cos} k_x ) s_y \tau_y - (t_{1p} \rm{cos} k_x + t_{2p} \rm{cos} k_y ) s_x \tau_y$,
$T=t_p - i t_d ( \rm{sin}k_x s_y + \rm{sin} k_y s_x ) \tau_y $,
$T_1 = m_2 (\rm{sin} k_x s_x + \rm{sin} k_y s_y ) \tau_x $. By fitting with the DFT results, we can get the corresponding parameters, $t_p=-0.0785, t_d=0.053, t_{1p}=-0.1331, t_{2p}=0.1597, m_1=0.0112, m_2=0.0006, \epsilon=0.3078$.
\end{widetext}
The surface band structures are calculated in an semi-infinite geometry by using the recursive Green's function method\cite{green} based on the previous tight-binding model.
\end{appendix}
\end{document}